**An introduction to network analysis for studies of medication use**

Mohsen Askar [1], Raphael Nozal Cañadas [2], Kristian Svendsen [1*]

[1] Department of Pharmacy, Faculty of Health Sciences, UiT The Arctic University of Norway

[2] Department of Informatics, Faculty of Science and Technology, UiT The Arctic University of Norway

[*]**Corresponding Author:** Kristian Svendsen

Post address: Faculty of Pharmacy, UiT Norges arktiske universitet, Postboks 6050 Langnes, 9037 Tromsø, Norway

E-mail address: kristian.svendsen@uit.no

**Abstract**

*Background*: Network Analysis (NA) is a method that has been used in various disciplines such as Social sciences and Ecology for decades. So far, NA has not been used extensively in studies of medication use. Only a handful of papers have used NA in Drug Prescription Networks (DPN). We provide an introduction to NA terminology alongside a guide to creating and extracting results from the medication networks.

*Objective:* To introduce the readers to NA as a tool to study medication use by demonstrating how to apply different NA measures on 3 generated medication networks.

*Methods*: We used the Norwegian Prescription Database (NorPD) to create a network that describes the co-medication in elderly persons in Norway on January 1, 2013. We used the Norwegian Electronic Prescription Support System (FEST) to create another network of severe drug-drug interactions (DDIs). Lastly, we created a network combining the two networks to show the actual use of drugs with severe DDIs. We used these networks to elucidate how to apply and interpret different network measures in medication networks.

*Results*: Interactive network graphs are made available online, Stata and R syntaxes are provided. Various useful network measures for medication networks were applied such as network topological features, modularity analysis and centrality measures. Edge lists data used to generate the networks are openly available for readers in an open data repository to explore and use.

*Conclusion*: We believe that NA can be a useful tool in medication use studies. We have provided information and hopefully inspiration for other researchers to use NA in their own projects. While network analyses are useful for exploring and discovering structures in medication use studies, it also has limitations. It can be challenging to interpret and it is not suitable for hypothesis testing.

*Keywords*: Network Analysis, Co-medication, Prescriptions, Drug interactions, Registries.

## Introduction

Studies in social pharmacy and pharmacoepidemiology often utilize highly complex data and require the use of sophisticated methods to discern important patterns. Data used for quantitative studies in social pharmacy and pharmacoepidemiology can be described as attribute data and relational data. Attribute data includes the characteristics of the studied objects (e.g. sex, age, medication use, sociodemographic information, etc.) while relational data contains the various relationships between subjects. The suitable way of studying attributes data is quantitative analyses, whereas, for relational data, Network Analysis (NA) is the appropriate approach [1]. The subjects studied in network analyses can take many different forms.

A network can be described as a graph that shows the interconnections between a set of actors. Each actor is represented by a *node* and each connection between these nodes is represented by an *edge* [2]. NA is a mathematical approach to study the relationships among nodes [3]. The mathematical background of NA are summarized elsewhere [4,5].

Network Analysis has its roots in many research disciplines [6]. Network analysis is used, among others, in social studies [7], ecological studies [8], genetics [9] and systems pharmacology [10].

As seen in figure 1, a network can be undirected (a and b) or directed (c and d). In a directed network, arrows show the direction of the relationship between nodes. In an undirected network, the relationship does not have a specific direction. The network edges can be weighted (b and d) or unweighted (a and c). In an unweighted network, the two nodes either have a relationship or not, while a weighed network considers the strength of the relationship.

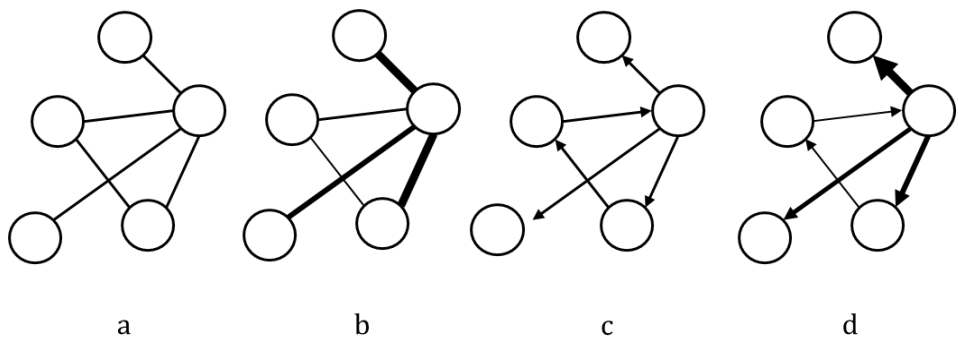

**Figure 1.** *Different types of networks, a) undirected and unweighted network, b) undirected and weighted network, c) directed and unweighted, and d) directed and weighted.*

*Use of Network Analysis in Public Health*

*Transmission networks* have been used to examine the risk of disease transmission by investigating the relations between the infected people and healthy ones [11–13]. Another form of transmission is the transmission of information. NA has been used to visualize the dissemination of public health information to different organizations and consumers. Some network characteristics reveal the pattern and the main actors contributing the most to information spread. Simulated networks can be used to suggest how to accelerate information spread [14]. An example of this type of networks, the diffusion of information among physicians regarding a new drug. The study showed that more socially integrated physicians introduced the drug months before corresponding isolated physicians [15]. NA was also used to study how health workers' professional and personal behavior impact health services [16,17].

*Drug Prescription Network (DPN)*

Pharmacoepidemiological studies of medications that are prescribed or dispensed is a relatively new application of NA. To our knowledge, Cavallo *et al.* were the first to study a drug prescription network in 2013. They used medications as the nodes and the number of patients being prescribed these medications as edges. They aimed mainly at describing the topology of the co-prescription network to demonstrate which drug classes are most co-prescribed. They also compared the male/female networks and networks from different age strata [18].

Bazzoni *et al.* were the first to use the term Drug Prescription Networks (DPN) in their paper published in 2015. They concluded that the DPNs are dense, highly clustered, modular and assortative. Density reflects frequent co-prescribing. Modularity suggested that the network could be subdivided into clusters. The study also showed that it is possible to highlight spatial and temporal changes by comparing different networks [19].

*Network Analysis terminology*

We organized the key measures that are useful in studies of medication use under 4 main categories: (1) Topology analysis (2) Modularity analysis (3) Network comparison (4) Bipartite networks [20]. (Figure 2)

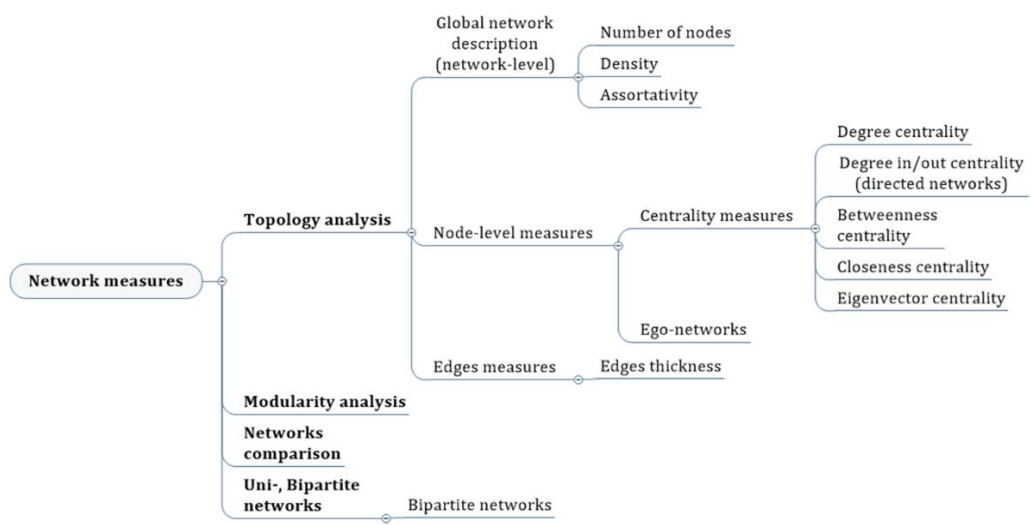

**Figure 2**. *Summarizing some of the Network Analysis measures that can be useful in the studies of medication use.*

1. <u>Topology analysis</u>

    Network topological features refer to a group of characteristics, which either describe the network as a whole (network-level) or define individual actors of the network (node-level). There are many topological measures and each of them gives information about a specific network attribute, which then may warrant further investigation.

a. *Global network description (network-level):* A group of measures that describe the network as a whole.

    - *Number of nodes*: the total number of drugs in the network. The network nodes can be grouped to show the number of drugs in each drug class. Different networks of different populations will have different distributions of drugs in the drug classes.

    - *Density:* the density of a network is the number of actual edges divided by the total number of edges that would exist if all the nodes in the network were connected. This potential number can be calculated by the formula below where $n$ is the number of nodes:

$$\frac{n \times (n-1)}{2}$$

The network density can be useful in terms of comparison between different networks that describe the same type of drug-drug relation.

*Assortativity:* a network is assortative when the nodes that share a similar trait tend to connect. This trait can be many characteristics such as the nodes' degree. In this case, the assortativity means that nodes with a high number of edges tend to connect. Assortativity can be examined in terms of other common characteristics between the nodes as well. Assortativity coefficient is measured using Pearson correlation. Assortativity coefficient is scaled between -1 and 1, where 1 is most assortative [21].

b. *Node-level measures*

Node-level measures describe the features of the different nodes across the network.

*Centrality measures*:

Centrality measures indicate the importance of the network nodes by assigning a score to each of them. There are many different centrality measures and each of them can be used to describe a specific type of importance. By comparing the different centrality measures of a node, we can understand the different ways a node is influential to the network. This paper will discuss 4 common types of centrality: degree, betweenness, closeness, and eigenvector centrality. The mathematical explanations of these measures are mentioned here [22,23].

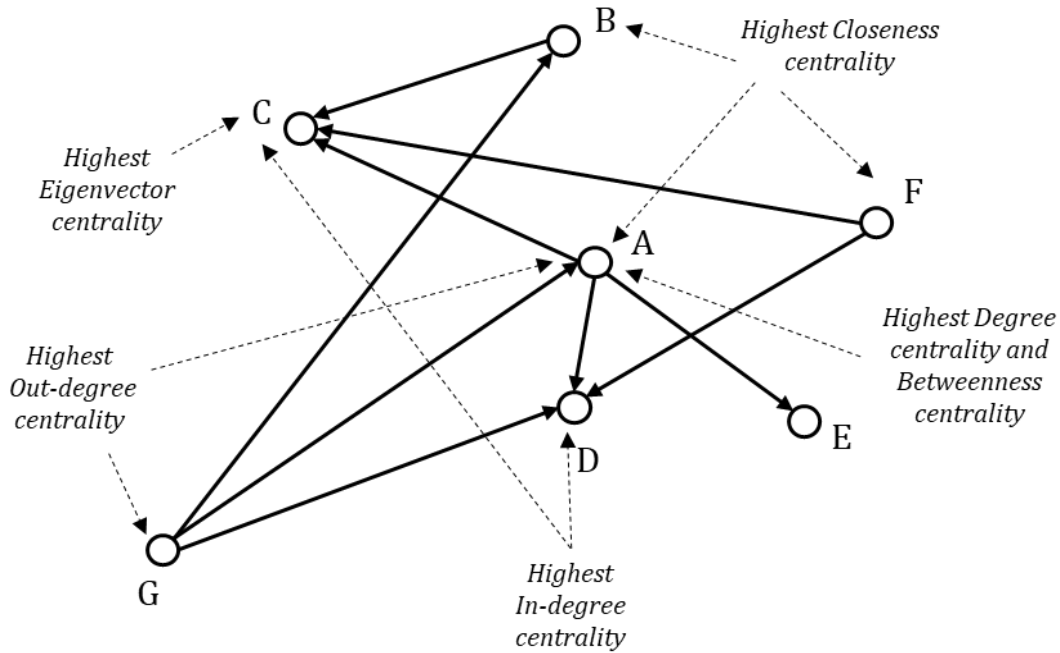

**Figure 3.** *Illustrating the different types of centrality. Node (A) represents the highest score of Degree and Betweenness centralities. The highest Eigenvector centrality score is assigned to the node (C). Nodes (A, B, F) have similar closeness Centrality. Nodes with the most in-degree edges are (C, D), while (A, G) have the most out-degree edges.*

*Degree centrality*

Degree centrality is the number of edges that are connected to a node. A higher score indicates that the node is connected to many other nodes. Node A in figure 3 has a degree score of 4. In a directed network, the degree is split into In-degree, which is the number of edges that direct to a node and Out-degree, which is the number of edges that originate from the node. In- and out- degrees will therefore show the directions of relationships in a network. In figure 3, nodes C and D have an in-degree score of 3, while nodes A and G have an out-degree score of 3.

*Betweenness centrality*

The betweenness centrality of a node indicates how many times this node was used to connect two other nodes by the shortest possible path. Increasing the number of shortest paths will increase the betweenness centrality score [22]. In figure 3, node A has the highest betweenness centrality score of 1.5.

*Closeness centrality*

It is a measure of the average distance between the node and all other nodes in the network. Nodes with the highest closeness score have the shortest distances to all other network nodes. The nodes A, B and F have the highest closeness centrality score of 1.

*Eigenvector centrality*

It is a measure of the importance of a node in a network based on the node's connections with other vital nodes. Relative scores are given to all nodes in the network based on the concept that connections to high-scored nodes give a higher score to the node than equal connections to low-scored nodes. In other words, a high eigenvector score means that a node is connected to many nodes, which themselves are connected to important nodes in the network and have high scores of eigenvector centrality. This means that a node with a high eigenvector centrality score is not necessarily connected to the highest number of nodes in the network but is connected to the nodes with a high number of edges [24]. Node C in figure 3 has the highest eigenvector centrality score of 1. Assigning the centrality of each node in the network may lead us to visualize the network from a single specific important node perspective; this is called an *Ego-network* and it visualizes the part of the network that has the node of interest and the nodes that are directly connected to it.

c. Edge-level measures

*Edge-thickness:* in a weighted network, the edge-thickness represents a quantitative measure of the strength of the connection between two nodes. This representation is unique for NA and can be used to study many research questions. We will show an example where the number of users that co-medicated a pair of medications are used to represent the edge-thickness. In this context, thicker edges represent more frequently used pairs of medications.

2. <u>Modularity analysis (Community detection)</u>

One key feature of the network structure is its modularity. A module is a group of nodes that have many connections between each other and few(er) connections to the other nodes in the network [25]. There are many techniques of community detection including density-based, centrality-based, partition-based and hierarchical clustering techniques [20,26,27]

3. <u>Network comparison</u>

It is possible to compare two or more networks to show the changes over time (temporal), between different areas (spatial), or between different groups of patients. These comparisons can be done by comparing the characteristics of the networks to highlight the differences in numbers and influences of the nodes. Another way to compare different networks is to subtract or divide the values of the edges between two networks. This will create edges representing the differences between the networks, see supplementary 4. By comparing many networks, dynamic graphs can be created showing the topological changes from a network to the next. Nodes will appear, disappear or change their locations as the dynamic graph moves through the different networks [28].

4. Bipartite networks

A network can be uni- or multipartite. We will only discuss uni- and bipartite networks. Unipartite networks have one set of nodes, while in bipartite networks the nodes belong to two disjointed sets (such as prescribers and patients). In a bipartite network, edges connect the nodes from different sets [29,30] (figure 4).

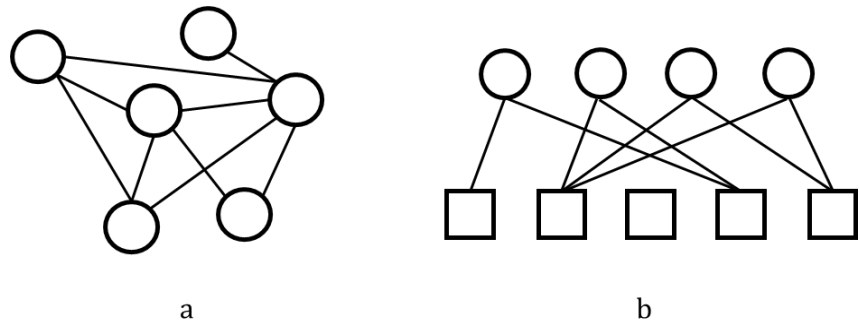

a                                          b

**Figure 4.** *(a) Unipartite network consisting of one type of nodes; (b) bipartite network consisting of two different types of nodes (circles and squares) in which the edges link between nodes from different types.*

The aim of this paper is to introduce the readers to NA as a tool to study medication use by demonstrating a practical real-life example of medication use in the elderly in Norway whenever it is possible, otherwise by giving an example from other studies

**Methods**

We created a network of co-medication in elderly persons in Norway. We also created a network describing the severe drug-drug interactions (DDIs). Finally, we generated a network with the actual use of drugs with severe DDIs by combining the previous two networks.

*Data sources*

*Co-medication network*

The dataset used comes from the Norwegian Prescription Database (NorPD). It covers all dispensed prescriptions to elderly persons (≥ 65 years) in Norway between 2012 and 2014. The NorPD collects data from all pharmacies in Norway and covers all outpatient dispensing for the entire Norwegian population. Details on the NorPD are published elsewhere [31]. In total, the dataset included 765,383 patients, 344,285 men (45%) and 421,098 women (55%) with 75 years as mean patient age. Edges in this network represent the number of patients who combined pairs of medications. In order to define the co-medication, we created treatment episodes using the Proportion of Days Covered (PDC) approach [32]. We assumed that patients used one Defined Daily Dose (DDD) [33] per day and added 20% to each prescription duration to account for imperfect adherence. We also allowed a medication-free gap of 14 days before ending a treatment episode and starting another. This means that if the patient exceeds 14 days without the medication, the treatment episode for this medication ends and a new episode starts if the patient picks up a new prescription. Finally, co-medication was defined as the overlapping drug treatment episodes at the index date, January 1, 2013.

For each pair of nodes (drugs), we summed up the number of co-medication occurrences (i.e. number of patients combining these two drugs) to create a weighted and undirected network.

We excluded the medications that have no defined DDD such as the medications for topical use, vaccines, and ophthalmologicals. In total, we excluded 357 medications (217 local and 140 systematic drugs). The co-medication network is shown here: [https://mohsenaskar.github.io/co-medication/network/](https://mohsenaskar.github.io/co-medication/network/). The network is searchable by substance name. Clicking on any node shows the ego-network of this node as well as some network measures.

*Severe drug-drug interactions network*

To create this network, we used a dataset derived from the Norwegian Electronic Prescription Support System (FEST). FEST is a national information service that provides common pharmaceutical data to the IT-systems that are involved in the drug prescribing process including systems used by physicians, hospitals and pharmacies [34]. Drug-drug interactions is a part of the FEST database. In FEST, the DDIs are divided into 3 categories; interactions that should be avoided (i.e. severe), interactions where precautions should be taken and interactions that do not require any action. Only severe DDIs were included in the study. There were 57,151 unique severe interactions. The edges in this network represent the presence of a severe interaction between the two nodes.

The network is undirected and unweighted. The severe DDIs network is shown here: [https://mohsenaskar.github.io/DDI/network/](https://mohsenaskar.github.io/DDI/network/)

*Combining co-medication and DDIs networks*

*Both DDI and co-medication network has drugs as nodes. When combining the two networks only edges that exists in both networks are included (only edges with any users combining the medications and where there is a severe DDI). The number of users for each edge from the co-medication network becomes the weight of the edges in the combined network.*

This network is shown here: [https://mohsenaskar.github.io/DDI-in-co-medication-network/network/](https://mohsenaskar.github.io/DDI-in-co-medication-network/network/)

*Preparing the data to create a network*

The data from the NorPD contains attributable data including a patient identity number, sex, year of birth, and data about each individual dispensed drug. To create a network, this data needed to be reshaped. The first step was to create a file with only medications that were used on the index date. Secondly, the file was aggregated such that an edge list was created. The edge list contains 2 variables defining the pairs of drugs and one variable with the number of users co-medicating with each pair of drugs. This edge list can be used by various software as described below. The process of data preparation is summed up in figure 5 and the edge list is openly available at the UiT The Arctic University of Norway open data repository here: https://dataverse.no/dataset.xhtml?persistentId=doi:10.18710/1OUTYI.The Stata syntax for creating the edge list is supplied in supplementary 2.

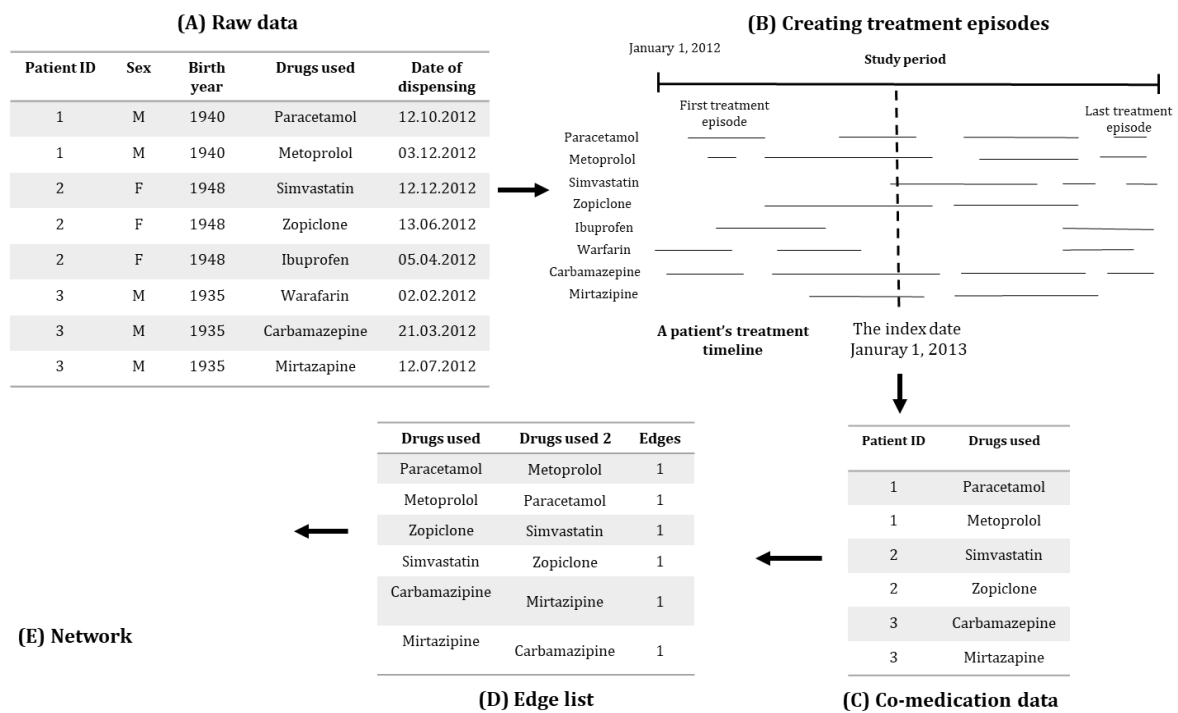

**Figure 5.** *Data preparation process. A) Raw clinical data in the long-form. B) Creating treatment episodes, a line indicates an episode. Gaps between lines indicate a medication-free period of more than 14 days. C) Data including only medications used at the index date D) Creating the edge list including three variables; two variables represent the nodes (Drug used, Drug used 2) and the third variable is the edge weight between the pair of nodes. E) Using the software to generate the network.*

*Software to use for network analysis*

There are many available tools to use for NA. We will focus on how to use the *nwcommands* package in *Stata* and the *igraph* package in *R* as well as visualization in *Gephi*. Other packages like "*igraph*" or "*NetworkX*" for *Python* are popular as well. All these packages can be used for visualizing and computing different network measures with differences in their integrated features and performance [35,36].

*Using Stata (nwcommands, nwANND)*

Using the edge list, *nwcommands* will create an adjacency matrix [37]. The adjacency matrix is a square matrix that contains the relationships between every pair of nodes in the network. The adjacency matrix can be saved as *Pajek* format that can be later imported and used by *Gephi*. In addition, *nwcommands* can display some network measures on both the network and node-level. *NwANND* is used for calculating the assortativity coefficient [38]. The syntax can be found in supplementary 2.

*Using R (igraph)*

Igraph *(https://igraph.org/)* is a library for creating and analyze graphs. It is widely used by network researchers to analyze graphs and networks. It is currently available for *C*, *C++*, *Python*, *R* and *Mathematica*.

One of the strengths of *igraph* is that it can be programmed with a high-level programming language and still be very efficient when handling large networks. In our *R* context, *igraph* integrates well with the visualization package (ggplot2) via the *ggraph* library.

Igraph uses an edge list and can link it with attribute data for each node as well. An example of network visualization using *Igraph* and *ggraph*, is given in Supplementary 2.

*Using Gephi*

*Gephi (https://gephi.org/users/)* is an open-source and free standalone software. The software can handle small to medium-sized networks (up to 150000 nodes). *Gephi* is user-friendly and requires no programming experience [36]. With many visualizing layouts and network measures, *Gephi* can provide a good starting point for the drug network study [39]. After importing the adjacency matrix to *Gephi*, we can process the network by applying different visualizing layouts, adding filters and colors. The structure of most drug networks can be complex and the unprocessed form of the network is often uninformative. By using different attributes (e.g. sex, modularity, etc.) the network can become more easily interpretable [28].

**Results**

We will present results from our networks using the same terms and order as in the introduction (figure 2). Table 1 provides the main topological features of the co-medication and the severe DDIs networks. The co-medication network is denser than the severe DDIs network, indicating that the drugs in the co-medication network are more connected. The assortativity coefficient shows that the co-medication network is non-assortative in terms of degree similarity, while the severe DDIs network is more assortative. Centrality measures in the co-medication network revealed that the same 5 drugs are the most central in all measures, while in the severe DDIs network; there is more variation in the top 5 drugs in each centrality measure. The results also showed that both networks are modular.

*Table 1. The topological measures of co-medication and severe DDIs networks*

| Outcome | Co-medication network | DDIs network | Indicates |
|---|---|---|---|
| **1. Topology analysis** | | | |
| **a) Network-level measures** | | | |
| **Number of nodes** | 762 | 1699 | The number of drugs present in the network. |
| **Number of edges** | 75052 | 57151 | Number of connections between the network nodes |
| **Density** | 0.26 | 0.04 | The extent of connections between the network's nodes |
| **Average degree** | 99 | 34 | The average number of connections that each node has. |
| **Assortativity co-efficient** | -0.26 | 0.4 | To what extent nodes with higher degrees tend to correlate. |
| **b) Node-level measures** | | | |
| *Centrality measures* | | | |
| **Nodes with the highest Degree centrality scores** | Acetylsalicylic acid Simvastatin Zopiclone Paracetamol Metoprolol | Typhoid vaccine Erythromycin Prikkperikum Clarithromycin Moxifloxacin | |
| **Nodes with the highest Betweenness centrality scores** | Acetylsalicylic acid Zopiclone Simvastatin Paracetamol Metoprolol | Typhoid vaccine Padeliporfin Hyperici herba (St John's-wort) Tuberculosis vaccine Ginkgo leaves | Combining these centrality measures can be used to assign the importance of each drug to the network. |
| **Nodes with the highest Closeness centrality scores** | Acetylsalicylic acid Simvastatin Zopiclone Paracetamol Metoprolol | Bromelains Telbivudine Peg interferon alfa-2a Diazepam Oxazepam | |
| **Nodes with highest Eigenvector centrality scores** | Acetylsalicylic acid Simvastatin Zopiclone Metoprolol Paracetamol | Typhoid vaccine Erythromycin Clarithromycin Chloramphenicol Moxifloxacin | |
| **c) Edge-level measures** | | | |
| **Average path length** | 1.77 | 3.09 | The average shortest path between two nodes. |
| **Thickest edge weight** | 82948 | 1 | For the weighted co-medication network the number reflects the highest number of patients co-medicating, This highlights clinically important combinations. |
| **Edges range** | 1-82948 | 0-1 | |
| **2. Modularity** | | | |
| **Modularity** | 0.088 | 0.54 | Indicates the presence of modules in the network. |
| **Number of modules (communities)** | 4 | 11 | |
| **Number of nodes in largest module** | 530 (module 0) | 372 (module 4) | |

Figure 6 shows that the majority of anatomical drug classes were assortative. This means that the drugs from the same anatomical group tend to be more co-prescribed. We also investigated the assortativity of the drugs on the pharmacological level (3rd level Anatomical Therapeutic Chemical classification) in supplementary 3.

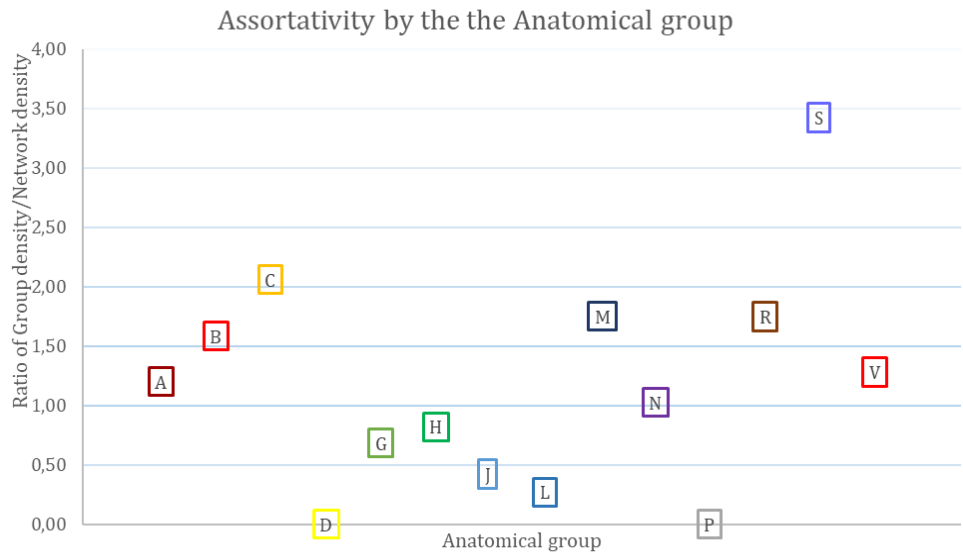

**Figure 6.** *Assortativity of network nodes in terms of similarity by the anatomical group. Squares above 1 represent a drug group with higher density than the general density of the network (0.26). The S (Sensory organs) anatomical group had the highest assortativity. D (Dermatologicals) and P (Antiparasitic products) groups had no edges because these drug classes were excluded from the study.*

Ego-networks as a measure can be seen by accessing the online networks we created and selecting individual nodes. The different network links can be found in the method section.

The top 10 edge weights for the severe DDIs in the co-medication network and co-medication only network are shown in tables 2 and 3 respectively. We see in table 2 that the number of patients using drugs causing severe DDIs are relatively low (less than 1000 users for all) while the most commonly co-medicated drugs seen in table 3 is much higher with acetylsalicylic Acid (aspirin) and simvastatin having around 83000 users representing almost 11% of the population.

***Table 2.*** *The top 10 clinically relevant severe DDIs in the co-medication network*

| | The severe DDI drug pair | | No. of patients co-medicating |
|---|---|---|---|
| 1 | Codeine and paracetamol | Tramadol | 855 |
| 2 | Esomeprazole | Clopidogrel | 823 |
| 3 | Simvastatin | Carbamazepine | 480 |
| 4 | Metoprolol | Paroxetine | 454 |
| 5 | Metoprolol | Verapamil | 380 |
| 6 | Lansoprazole | Clopidogrel | 308 |
| 7 | Diclofenac | Ibuprofen | 305 |
| 8 | Diazepam | Oxazepam | 300 |
| 9 | Carbamazepine | Zopiclone | 280 |
| 10 | Omeprazole | Clopidogrel | 277 |

***Table 3.*** *The top 10 combined drugs in the co-medication network*

| | Most combined drugs | | No. of patients co-medicated |
|---|---|---|---|
| 1 | Acetylsalicylic acid | Simvastatin | 82948 |
| 2 | Acetylsalicylic acid | Metoprolol | 52577 |
| 3 | Acetylsalicylic acid | Atorvastatin | 42753 |
| 4 | Metoprolol | Simvastatin | 36792 |
| 5 | Acetylsalicylic acid | Amlodipine | 32628 |
| 6 | Acetylsalicylic acid | Zopiclone | 29173 |
| 7 | Amlodipine | Simvastatin | 22554 |
| 8 | Acetylsalicylic acid | Ramipril | 19660 |
| 9 | Simvastatin | Zopiclone | 18845 |
| 10 | Metformin | Acetylsalicylic acid | 18507 |

*Modularity analysis*

We found 4 modules in the co-medication network and 11 modules in the severe DDI network. For the co-medication network, there was one large community and 3 other smaller communities.. Nervous and Respiratory system groups (N- and R- groups) drugs are just found in module 0, while Cardiac-, Alimentary-, Blood groups(C-, A-, B- groups) are common groups between modules 1 and 2, but with considering the number of users in each module we can locate in which module these ATC groups represent the most importance. Drugs used for diabetes, (A10) group, present only in module 2. The complete tables of modules are listed in supplementary 1. For the severe DDIs network, the modules found are shown below in figure 7.

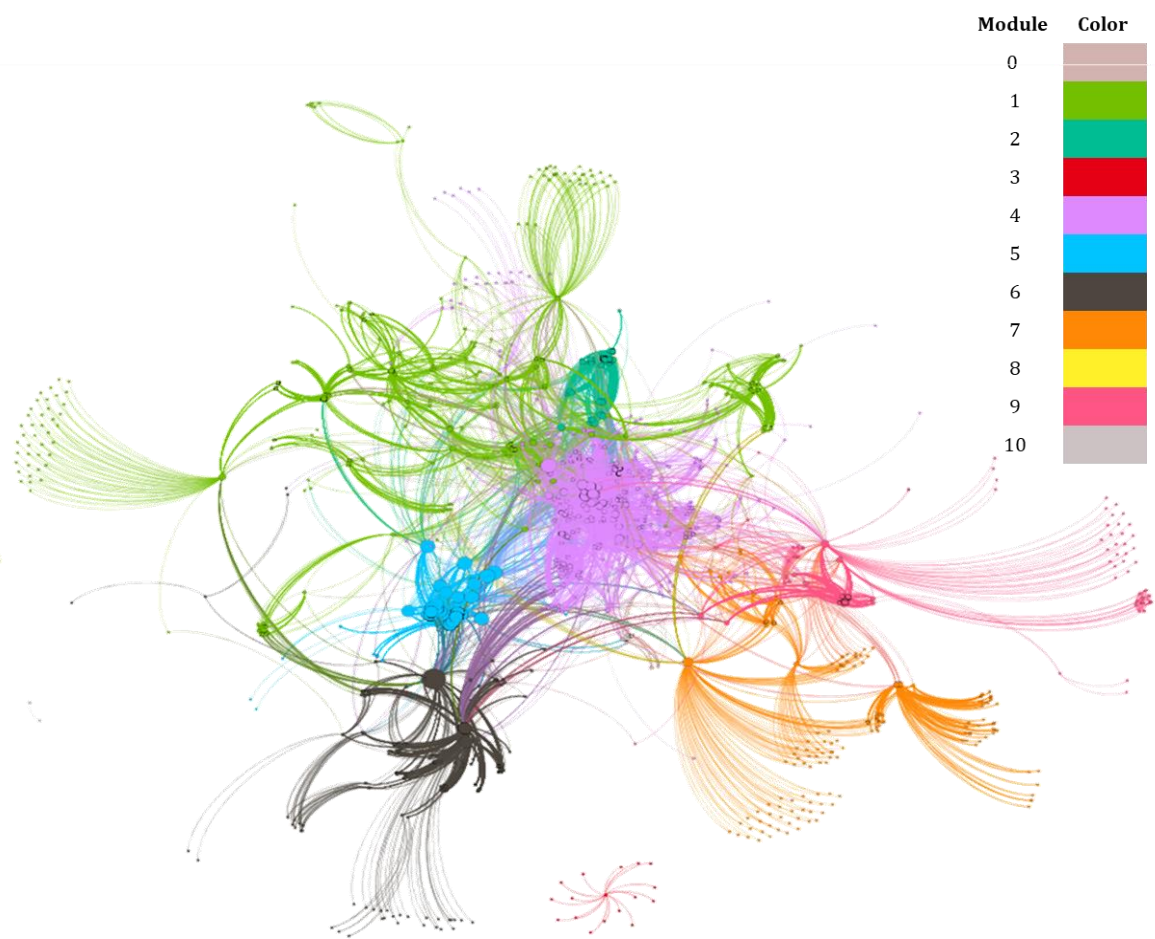

**Figure 6.** *The 11 modules that were detected in the severe DDIs network. Different colors indicate different modules.*

**Discussion**

A Network visualizes the relationships of a dataset in one graph. This unique ability of data representation is combined with many measures that are helpful for many research disciplines. A starting point for generating any network is to select the nodes and define the edges. A precise definition of the edges allows the researcher to extract the correct information. NA is a well-suited approach to study complex systems. Although the approach has been widely used in many fields of research, only a few studies studied the drug-relations in a network [18,19].

Our results show that many network outcomes can be useful in the studies of medication use. Moreover, some results are unique measures that only NA can perform such as edge measure and modules detection. Employing centrality measures in the drug study introduce an opportunity to observe the influence of the different drugs in the drug-network. Determining this influence can be useful for clinicians and decision-makers.

After generating a network, some topological features have to be reported first to get a general idea about the network content and its basic characteristics. Network-level measures such as assortativity and density reveal many clues for further investigation. Centrality measures show how influential each node is in the network. It is possible to have high centrality of one type and a low of another for the same node [8]. In order to study the importance of the nodes, it is necessary to use more than one measure of centrality. Recent studies suggest using centrality measures as an alternative approach for variables selection. Lutz *et al* used the centrality measures to identify 4 additional variables contributing to the predicting of treatment dropout in patients with anxiety disorders [40]. Valenzuela *et al.* described a methodology based on degree and centrality measures to obtain the most representative variables for predicting successful aging [41]. These approaches are interesting and represent an alternative method to the other variable selection methods. Edge-level measures are the core of the networks and the principal for many network measures.

Modularity analysis exposes the network structure. This measure is believed to introduce special importance in the drug study. Bazzoni *et al.* found the DPN to be modular [19], which is consistent with what we found in our networks. Further investigation is needed to assess the underlying patterns in the modules we found in our co-medication network. Modules can be interpreted as clusters of patients with similar diagnoses using the same medications. In our initial analysis of modularity, we identified 4 modules, further work could be done to identify smaller groups by detecting the sub-clusters inside each module. Modules in the DDI network could be connected to pharmacological data to see the importance of pharmacokinetic interactions through systems such as the cytochrome P450 system. We have not explored this but there is a great potential in using modularity analyses in order to understand networks.

Comparing different networks can reveal the change in patterns over time, place and different populations. Networks that describe the relation between drug-use and morbidities for a patient or a group of similar patients over time may identify the development of co-morbidities and drug use.

Bipartite networks provide a variety of possibilities to study many situations in which drugs are involved with other network actors such as physicians or diseases. Dasgupta & Chawla created a bipartite drug-disease network to study the interactions between drugs and co-morbidities [42]. Hu *et al.* studied the prescribing of some opioids by creating a bipartite network of patients and prescribers and using the network to analyze the relationship between patients and prescribers and detect the "doctor shopping" and suspicious network nodes. A redrawn example from this study is shown in figure 8.

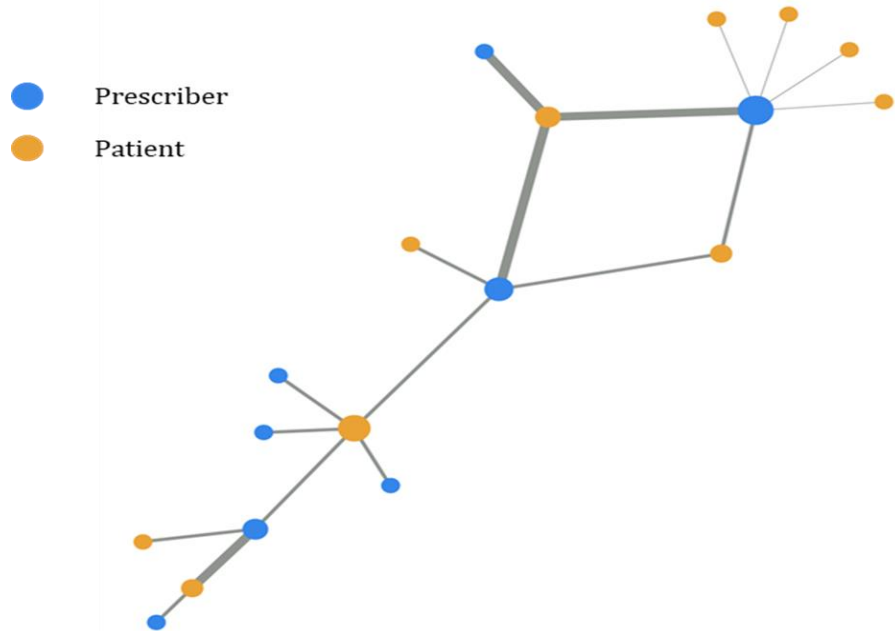

**Figure 8.** *An example of a bipartite network representing a sub-graph of two types of nodes (i.e. prescribers and patients) linked by the number of Fentanyl® patches prescriptions. The bigger nodes indicate more number of connections. The thicker edges indicate a higher number of prescriptions. (Redrawn from "Network analysis and visualization of opioid prescribing data" [43]).*

Our study has some limitations. As we used the DDD to outline the treatment episodes, we excluded the medications that have no defined DDD. This reduced the represented co-medication in our networks to the actual co-medication at the index date.

NA also has some important limitations. As a tool, it can be used to explore data, to find unusual structures, group nodes together and find unusual individual nodes. However, it can be hard to interpret results from NA and it is only suited for hypothesis generation. It also cannot explore many sets of relationships between variables at the same time as well as determining causal relationships. For such research questions, other hypothesis testing methodologies will be more needed. However, in research focused on exploration, NA can be a valuable tool.

**Conclusion**

The main purpose of this paper was to demystify the NA as a method. We have explained the terminology of network analyses and showed, with examples, how network analyses can be used for hypothesis generation. The online links to our networks visualize the data much better than a static picture can and we hope that we have provided enough information, and inspiration, to explore how you can use NA on your own data. We are confident that the future will see many new applications of NA and interesting results for researchers in social pharmacy and pharmacoepidemiology.

**CRediT authorship contribution statement**

**Mohsen Askar:** Conceptualization, Methodology, Software, Data Curation, Writing – Original Draft, Writing – Review & Editing, Visualization. **Kristian Svendsen:** Conceptualization, Software, Writing – Review & Editing, Supervision. **Raphael N. Cañadas:** Software, Writing – Review & Editing.

**Funding**

This research did not receive any specific grant from funding agencies in the public, commercial, or not-for-profit sectors.

**Declaration of competing interest**

The authors declare that they have no conflicts of interest related to this study.

**Acknowledgments**

We thank Lars Småbrekke and Elin Lehnbom for their useful feedback on the manuscript.

**Supplementary material**

Supplementary material associated with this article can be found in the online version at ..

**Supplementary 1:**

**Modularity in the co-medication network**

We found 4 modules in the co-medication network. The number of medications (nodes) under each module is shown in table 1. By collapsing the ATCs to the therapeutic level (2nd level), we could group the ATC codes under their corresponding therapeutic group and reduce the number of studied objects in each module (table 2).

We integrated the number of users of each therapeutic group to assign the importance of the therapeutic groups to the module if the therapeutic group was common between more than one module. Another step that could have been done is to detect the sub-clusters inside the complex module (module 0) to see if this could further explain the underlying pattern in this module. Moreover, diagnosis can be related to the therapeutic groups in each module to recognize the comorbidities that underlie each module.

*Table 1.1 The four modules found in the co-medication network*

| Module | Number of nodes | Percent |
|---|---|---|
| 0 | 530 | 68 % |
| 1 | 49 | 24 % |
| 2 | 167 | 6 % |
| 3 | 16 | 2 % |

*Table 1.2.* The four modularity classes obtained from our co-medication network. We collapsed ATCs to the third level to simplify the patterns. Some ATC groups are unique for the module, and others are common between more than one. It is important to take the number of users into consideration when trying to understand each pattern.

| Group | No. of users | Therapeutic indications |
|---|---|---|
| colspan="3" | **Modularity class 0** | |
| N05 * | 149142 | HYPNOTICS AND SEDATIVES |
| R03 * | 125913 | DRUGS FOR OBSTRUCTIVE AIRWAY DISEASES |
| A02 | 107657 | ACID DISORDERS |
| N02 * | 87051 | ANTIMIGRAINE PREPARATIONS |
| N06 * | 78403 | PSYCHOANALEPTICS |
| M01 * | 60442 | ANTIINFLAMMATORY AND ANTIRHEUMATIC PRODUCTS |
| B03 | 57100 | ANTIANEMIC PREPARATIONS |
| H03 * | 56354 | IODINE THERAPY |
| R06 * | 53993 | ANTIHISTAMINES FOR SYSTEMIC USE |
| J01 * | 40218 | ANTIBACTERIALS FOR SYSTEMIC USE |
| M05 | 35792 | Bone diseases |
| H02 * | 35343 | CORTICOSTEROIDS FOR SYSTEMIC USE |
| G04 | 23622 | UROLOGICALS (prostatic hypertrophy) |
| R01 * | 21306 | NASAL PREPARATIONS |
| G03 * | 20842 | SEX HORMONES |
| N03 * | 17032 | ANTIEPILEPTICS |
| R05 * | 16266 | COUGH AND COLD PREPARATIONS |
| C03 | 14703 | DIURETICS |
| L04 | 12827 | IMMUNOSUPPRESSANTS |
| N04 * | 10016 | ANTI-PARKINSON DRUGS |
| A06 | 7932 | DRUGS FOR CONSTIPATION |
| A07 | 6774 | ANTIDIARRHEALS, INTESTINAL ANTIINFLAMMATORY/ANTIINFECTIVE AGENTS |
| L02 | 5189 | ENDOCRINE THERAPY (HORMONS RELATED) |
| A11 | 5113 | VITAMINS |
| B01 | 4793 | ANTITHROMBOTIC AGENTS |
| A03 | 4159 | DRUGS FOR FUNCTIONAL GASTROINTESTINAL DISORDERS |
| C07 | 3349 | BETA BLOCKING AGENTS |
| C10 | 3309 | LIPID MODIFYING AGENTS |
| N07 * | 1823 | OTHER NERVOUS SYSTEM DRUGS |
| A12 | 1631 | MINERAL SUPPLEMENTS |
| A09 | 1519 | DIGESTIVES, INCL. ENZYMES |
| P01 * | 1342 | ANTIPROTOZOALS |
| D01 | 1262 | ANTIFUNGALS FOR DERMATOLOGICAL USE |
| H01 * | 804 | PITUITARY AND HYPOTHALAMIC HORMONES AND ANALOGUES |
| M03 | 742 | MUSCLE RELAXANTS |
| A04 | 735 | ANTIEMETICS AND ANTINAUSEANTS |
| A05 | 595 | BILE AND LIVER THERAPY |
| A08 | 516 | ANTIOBESITY PREPARATIONS, EXCL. DIET PRODUCTS |
| L01 | 476 | ANTINEOPLASTIC AGENTS |
| J05 | 392 | ANTIVIRALS FOR SYSTEMIC USE |
| **L03** | 288 | IMMUNOSTIMULANTS |
| **J04** | 269 | ANTIMYCOBACTERIALS |
| **D05** | 247 | ANTIPSORIATICS |
| **J02** | 188 | ANTIMYCOTICS FOR SYSTEMIC USE |
| **A01** | 185 | STOMATOLOGICAL PREPARATIONS |
| **H05** | 131 | CALCIUM HOMEOSTASIS |
| **G02** | 115 | OTHER GYNECOLOGICALS |
| **B02** | 113 | ANTIHEMORRHAGICS |
| | **Modularity class 1** | |
| **A11** | 1635 | VITAMINS |
| **C03** | 70736 | DIURETICS |
| **B01** | 52433 | ANTITHROMBOTIC AGENTS |
| **M04** * | 15650 | ANTIGOUT PREPARATIONS |
| **C01** | 15616 | CARDIAC THERAPY |
| **C07** | 13343 | BETA BLOCKING AGENTS |
| **C08** | 7445 | CALCIUM CHANNEL BLOCKERS |
| **A12** | 5858 | MINERAL SUPPLEMENTS |
| **B03** | 1282 | ANTIANEMIC PREPARATIONS |
| **V03** | 487 | ALL OTHER THERAPEUTIC PRODUCTS |
| **H05** | 240 | CALCIUM HOMEOSTASIS |
| **A02** | 194 | DRUGS FOR ACID RELATED DISORDERS |
| **C02** | 163 | ANTIHYPERTENSIVES |
| **L04** | 161 | IMMUNOSUPPRESSANTS |
| | **Modularity class 2** | |
| **C09** | 293008 | AGENTS ACTING ON THE RENIN-ANGIOTENSIN SYSTEM |
| **B01** | 260482 | ANTITHROMBOTIC AGENTS |
| **C10** | 258050 | LIPID MODIFYING AGENTS |
| **C07** | 128165 | BETA BLOCKING AGENTS |
| **C08** | 126834 | CALCIUM CHANNEL BLOCKERS |
| **A10** * | 96241 | DRUGS USED IN DIABETES |
| **S01** * | 67468 | OPHTHALMOLOGICALS |
| **G04** | 45302 | UROLOGICALS (prostatic hypertrophy) |
| **C01** | 26825 | CARDIAC THERAPY |
| **C03** | 22963 | DIURETICS |
| **L02** | 11148 | ENDOCRINE THERAPY |
| **C02** | 10313 | ANTIHYPERTENSIVES |
| **L01** | 708 | ANTINEOPLASTIC AGENTS |
| **C04** * | 467 | PERIPHERAL VASODILATORS |
| **A07** | 171 | ANTIDIARRHEALS, INTESTINAL ANTIINFLAMMATORY/ANTIINFECTIVE AGENTS |
| | **Modularity class 3** | |
| **J05** * | 237 | ANTIVIRALS FOR SYSTEMIC USE |

(*) unique drug group for this module

**Supplementary 2**

**Stata and R syntaxes**

The STATA syntax for creating an edge list of co-medicated drugs and R syntax for some examples of networks generation are uploaded here: https://github.com/MohsenAskar/NA-in-medication-study

Instructions for installing STATA are available at https://www.stata.com/ . Instructions for installing nwcommands package for network analysis are available at https://nwcommands.wordpress.com/installation/ and for installing NwANND are available here https://ideas.repec.org/c/boc/bocode/s458261.html . Tutorials for the use of nwcommands are here https://nwcommands.wordpress.com/tutorials-and-slides/.

Installing of R is available here https://cran.r-project.org/doc/manuals/r-release/R-admin.html. Instructions for installing igraph are here https://cran.r-project.org/web/packages/igraph/readme/README.html. More on igraph is here https://igraph.org/.

Gephi installing instructions are available here https://gephi.org/ and tutorials are available here https://gephi.org/users/.

# Supplementary 3

## Assortativity

We examined the assortativity of the drugs from the same anatomical group to connect to each other. We calculated the ratio between the densities in-between drugs from the same anatomical class to the density of the general network (0.26). If the ratio was higher than 1, this means a higher density between these drugs than the network density and indicates a tendency to correlate. This means that the drugs from the same anatomical group tend to be more co-prescribed.

**Assortativity of the anatomical classes (by ATC codes similarity):**

*Table 3.1: Assortativity of anatomical groups by ATCs similarity*

| Group | No. of ATCs in each group | Potential no. of edges * | Actual no. of edges | Density between ATCs of the same anatomical group ** | Ratio Density (group/network density) |
|---|---|---|---|---|---|
| A | 108 | 5778 | 1802 | 0.312 | 1.20 |
| B | 35 | 595 | 245 | 0.412 | 1.58 |
| C | 98 | 4753 | 2544 | 0.535 | 2.06 |
| D | 6 | 15 | 0 | 0.000 | 0.00 |
| G | 49 | 1176 | 209 | 0.178 | 0.68 |
| H | 27 | 351 | 75 | 0.214 | 0.82 |
| J | 75 | 2775 | 305 | 0.110 | 0.42 |
| L | 70 | 2415 | 170 | 0.070 | 0.27 |
| M | 32 | 496 | 226 | 0.456 | 1.75 |
| N | 166 | 13695 | 3653 | 0.267 | 1.03 |
| P | 9 | 36 | 0 | 0.000 | 0.00 |
| R | 69 | 2346 | 1066 | 0.454 | 1.75 |
| S | 14 | 91 | 81 | 0.890 | 3.42 |
| V | 4 | 6 | 2 | 0.333 | 1.28 |

*(*) Expected number of edges if all nodes were connected. Calculated as N*(N-1)/2*

*(**) Calculated as actual number edges/theoretical number of edges*

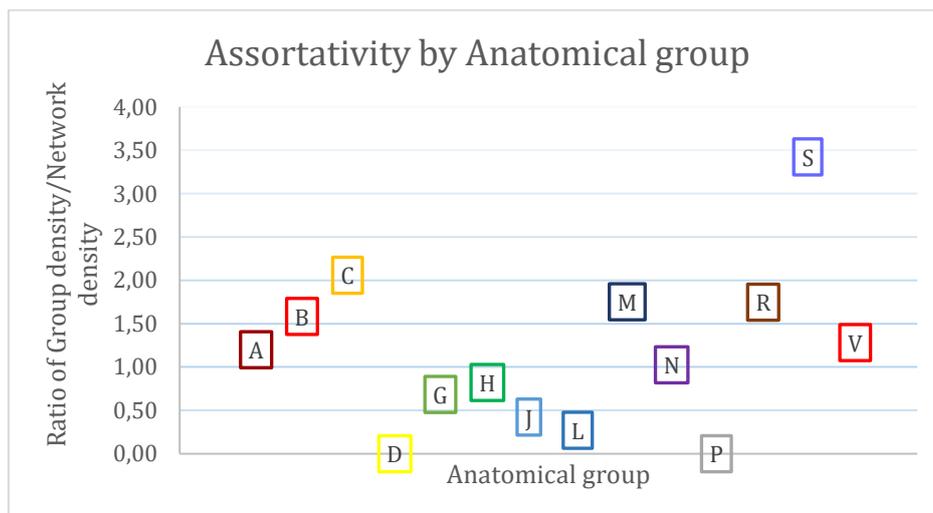

***Figure 3.1*** *Assortativity of network nodes in terms of similarity by the anatomical group. Groups that lie over 1 represent a higher density between their ATC codes than the density of the network (0.26). D-, P- groups have no edges between themselves mainly because of the exclusion criteria of the study, see methods chapter.*

***Table 3.2.*** *Assortativity of pharmacological groups (3rd level ATC codes) by ATC codes similarity*

| Anatomical Group | Ratio Density (anatomical group/network density) | pharmacological groups | Ratio Density (pharmacological group/network density) |
|---|---|---|---|
| A | 1.199510078 | A01A | 3.846154 |
| | | A02B | 2.393162 |
| | | A03A | 3.846154 |
| | | A04A | 0.6410257 |
| | | A06A | 1.611722 |
| | | A07D | 2.564103 |
| | | A07E | 1.785714 |
| | | A09A | 3.846154 |
| | | A10A | 3.418803 |
| | | A10B | 2.388664 |
| | | A11C | 2.307692 |
| | | A12B | 3.846154 |
| | | A12C | 2.564103 |
| B | 1.583710407 | B01A | 1,518219 |
| | | B03A | 2.564103 |
| | | B03B | 3.461538 |
| | | B03X | 2.564103 |
| C | 2.058618848 | C01A | 3.846154 |
| | | C01B | 1.025641 |
| | | C01D | 3.846154 |
| | | C02A | 1.282051 |
| | | C03A | 3.846154 |
| | | C03C | 1.282051 |
| | | C03D | 1.282051 |
| | | C07A | 2.029915 |
| | | C08C | 3.076923 |
| | | C08D | 3.846154 |
| | | C09A | 2.692308 |
| | | C09B | 2.564103 |
| | | C09C | 2.930403 |
| | | C09D | 2.735043 |
| | | C10A | 2.262444 |
| G | 0.683542648 | G03C | 1.923077 |
| | | G03F | 2.564103 |
| | | G04B | 1.648352 |
| | | G04C | 3.846154 |
| H | 0.821827745 | H01C | 3.846154 |
| | | H02A | 1.111111 |
| | | H03A | 3.846154 |
| | | H03B | 3.846154 |
| | | H05B | 1.282051 |
| J | 0.422730423 | J01A | 1.282051 |
| | | J01C | 1.068376 |
| | | J01E | 3.846154 |

|   |   | J01F | 1.923077 |
|---|---|------|----------|
|   |   | J01X | 0.7692308 |
|   |   | J02A | 0.3846154 |
|   |   | J04A | 1.923077 |
|   |   | J05A | 0.4782255 |
| L | 0,270743749 | L02A | 1,282051 |
|   |   | L02B | 1,367521 |
|   |   | L03A | 0,1709402 |
|   |   | L04A | 0.7894737 |
| M | 1.75248139 | M01A | 2.600733 |
|   |   | M03B | 1.282051 |
|   |   | M04A | 3.205128 |
|   |   | M05B | 0.8241758 |
| N | 1.025921869 | N02A | 1.826923 |
|   |   | N02B | 1.025641 |
|   |   | N02C | 1.623932 |
|   |   | N03A | 1.758242 |
|   |   | N04B | 2.797203 |
|   |   | N05A | 1.611432 |
|   |   | N05B | 1.025641 |
|   |   | N05C | 1.709402 |
|   |   | N06A | 1.538462 |
|   |   | N06D | 3.846154 |
|   |   | N07A | 1.282051 |
|   |   | N07B | 0.4273504 |
| R | 1.747655584 | R01A | 1.141167 |
|   |   | R03A | 2.991453 |
|   |   | R03B | 2.393162 |
|   |   | R03D | 2.307692 |
|   |   | R05C | 2.564103 |
|   |   | R05D | 2.307692 |
|   |   | R06A | 1.245421 |
| S | 3.423499577 | S01E | 3.42 |
| V | 1.282051282 | V03A | 1.282051 |

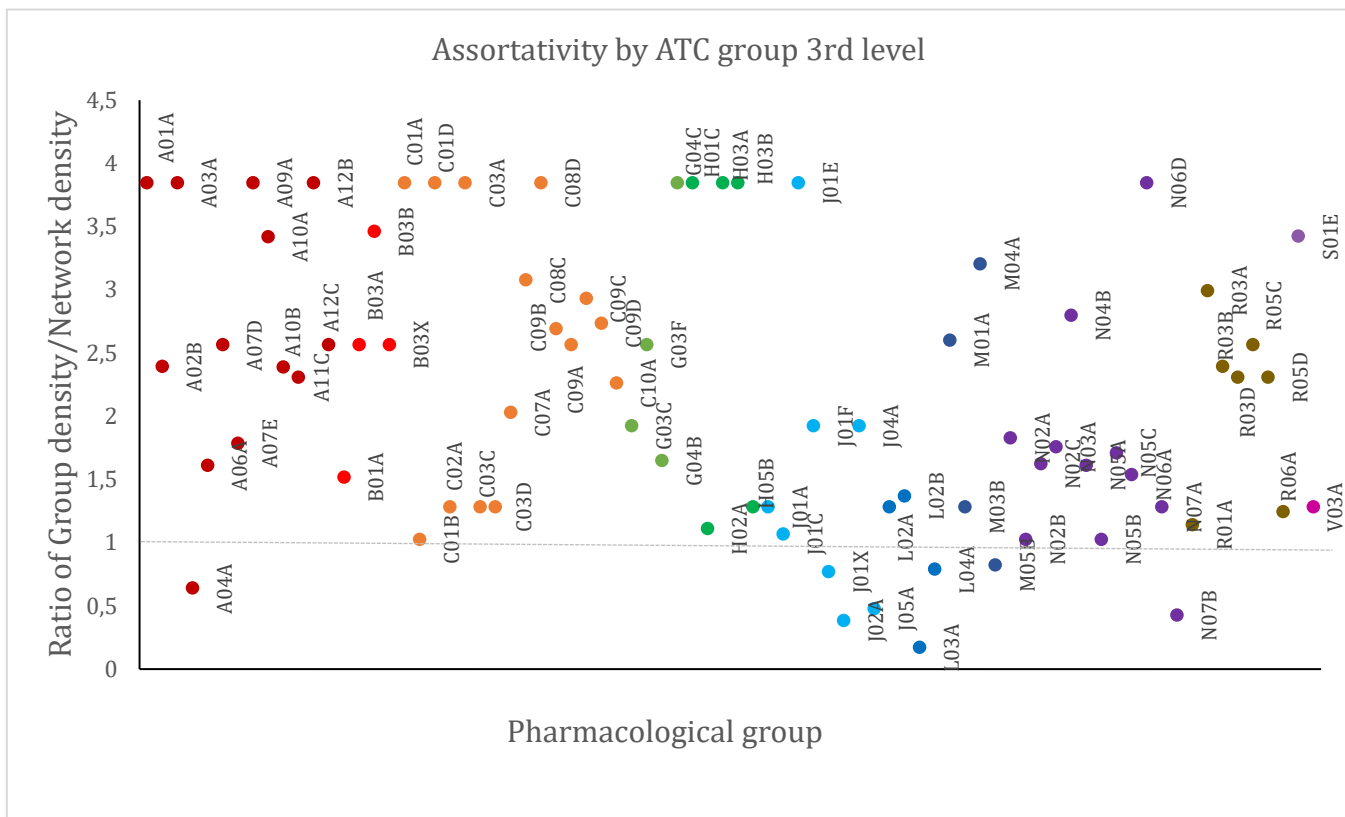

***Figure 3.2.*** *Assortativity of the different pharmacological groups. Groups above 1 indicate a higher density between them than the density of the general network. Most of the groups lie above 1 in the figure indicating a tendency to be co-medicated together.*

**Supplementary 4**

**Network comparison**

As an example of temporal comparison, we generated another co-medication network representing the medication-use on January 1, 2014, to compare it with our original co-medication network that represents the medication-use on January 1, 2013.

We generated a new network that shows the difference between 2013 and 2014 co-medication networks. We did this by dividing the weight of each edge (number of patients who used these pairs of medications) of the 2013 network by the weight of the same edges in the 2014 network. This resulted in (1) unmatched edges that are unique for each network and (2) new weights of the matched networks that represent the ratio of users of the common edges in the 2013 network to the 2014 network. For example, if we have a pair of medicines that were used by 20 patients in the 2013 network and 10 patients in the 2014 network the new edge weight in the generated network will equal to 2, which means that this combination was used twice as many times in 2013 compared to 2014. In total, 84% of the edges were common between the two networks, while 16% unique edges in both. Only 19% of drug-drug combinations remain constant in both years, while 31% of overall combinations are higher used in 2014 than in 2013 and 50% of them are higher used in 2013 than in 2014 (table 4.1).

*Table 4.1 Drug-combination frequency comparison (2013-2014)*

|  | Frequency | Percent |
|---|---|---|
| **Lower combination frequency in 2013 than 2014** | 19381 | 31 % |
| **No change** | 11861 | 19 % |
| **Higher combination frequency in 2013 than 2014** | 31567 | 50 % |
| **Total** | 62809 | 100 % |

Table 4.2 represents the most used drug pairs in 2013 compared to 2014 and vice versa in table 4.3. There are many reasons of which why these combinations vary, including withdrawal of some medicines from the drug market e.g. digitoxin, or approving of others e.g. dapagliflozin and apixaban. More significant differences are expected to take place if the time gap between the two DPNs is longer.

*Table 4.2 Top combinations used with higher frequency in 2013 compared to 2014*

| ATC 1 | Medication 1 | ATC 2 | Medication 2 | Ratio users 2013/2014 |
|---|---|---|---|---|
| C01AA04 | Digitoxin | L04AX03 | Methotrexate | 21 |
| B01AB04 | Deltaparin (Fragmin ®) | M01AB55 | Diclofenac combinations | 19 |
| C01AA04 | Digitoxin | R03BA02 | Budesonide | 17 |
| A11CC01 | Ergocalciferol (AFI-D2 forte ®) | J01CA04 | Amoxicillin | 16 |
| C01AA04 | Digitoxin | C09AA01 | Captopril | 16 |
| M01AB05 | Diclofenac | N02AB01 | Ketobemidone (Ketorax ®) | 16 |
| A04AA01 | Ondansetron (Zofran ®) | J01EE01 | Sulfamethoxazole and Trimethoprim | 15 |
| C01AA04 | Digitoxin | N06AA09 | Amitriptyline | 15 |
| C01DA08 | Isosorbide dinitrate | J01CE02 | Phenoxymethylpenicillin | 15 |
| C03EA01 | HCT/Pot. sparing agents | P01AB01 | Metronidazole | 15 |
| G04CA01 | Alfuzosin (Xatral ®) | G04CA02 | Tamsulosin | 15 |

*Table 4.3 Top combinations used with higher frequency in 2014 compared to 2013*

| ATC 1 | Medication 1 | ATC 2 | Medication 2 | Ratio users 2014/2013 |
|---|---|---|---|---|
| A10BK01 | Dapagliflozin (Forxiga ®) | C10AA05 | Atorvastatin | 143 |
| B01AF02 | Apixaban (Xarelto ®) | C07AB07 | Bisoprolol | 102 |
| C09DA04 | Irebsartan/HCT (CoAprovel®) | G04BD12 | Mirabegron (Betmiga ®) | 74 |
| A10BA02 | Metformin | A10BK01 | Dapagliflozin (Forxiga ®) | 73 |
| A10BK01 | Dapagliflozin (Forxiga ®) | B01AC06 | Acetyl salicylic acid | 71 |
| B01AF02 | Apixaban (Xarelto ®) | C01DA14 | Isosorbide mononitrate | 71 |
| A10BK01 | Dapagliflozin (Forxiga ®) | C10AA01 | Simvastatin | 70 |
| A02BC02 | Pantoprazole | B01AF02 | Apixaban (Xarelto ®) | 68 |
| B01AF02 | Apixaban (Xarelto ®) | C09CA01 | Losartan | 64 |
| A10BD08 | Metformin/Vildagliptin (Eucreas®) | A10BK01 | Dapagliflozin (Forxiga ®) | 62 |